\documentstyle[12pt,aaspp4,flushrt]{article}

\catcode`\@=11 
\def\@versim#1#2{\vcenter{\offinterlineskip
        \ialign{$\m@th#1\hfil##\hfil$\crcr#2\crcr\sim\crcr } }}
\newcommand{\beq}{\begin{equation}}
\newcommand{\eeq}{\end{equation}}
\def\lsim{\mathrel{\mathpalette\@versim<}}
\def\gsim{\mathrel{\mathpalette\@versim>}}

\def\pp{p_\perp}
\def\ppar{p_\parallel}
\def\kpar{k_\parallel}
\def\qp{q_\perp}
\def\qpar{q_\parallel}
\def\dpp{\delta p_\perp}
\def\dppar{\delta p_\parallel}
\def\drho{\delta \rho}
\def\db{\delta B}
\def\kpar{k_\parallel}
\def\kp{k_\perp}

\begin{document}

\title{The Magnetorotational Instability in a Collisionless Plasma}
\author{Eliot Quataert,\altaffilmark{1} William Dorland,\altaffilmark{2,3} and Greg Hammett\altaffilmark{2,4}} 
\altaffiltext{1}{UC Berkeley, Astronomy Department, 601 Campbell Hall, Berkeley, CA 94720; eliot@astron.berkeley.edu} 
\altaffiltext{2}{Imperial College, Blackett Laboratory, Prince Consort Rd, London SW7 2BW, UK}
\altaffiltext{3}{bdorland@kendall.umd.edu}
\altaffiltext{4}{Permanent Address: Princeton Plasma Physics Laboratory, P. O. Box 451, Princeton, NJ, 08543; hammett@princeton.edu}
\medskip

\begin{abstract}

We consider the linear axisymmetric stability of a differentially
rotating collisionless plasma in the presence of a weak magnetic
field; we restrict our analysis to wavelengths much larger than the
proton Larmor radius.  This is the kinetic version of the
magnetorotational instability explored extensively as mechanism for
magnetic field amplification and angular momentum transport in
accretion disks.  The kinetic calculation is appropriate for hot
accretion flows onto compact objects and for the growth of very weak
magnetic fields, where the collisional mean free path is larger than
the wavelength of the unstable modes.  We show that the kinetic
instability criterion is the same as in MHD, namely that the angular
velocity decrease outwards.  However, nearly every mode has a linear
kinetic growth rate that differs from its MHD counterpart.  The
kinetic growth rates also depend explicitly on $\beta$, i.e., on the
ratio of the gas pressure to the pressure of the seed magnetic field.
For $\beta \sim 1$ the kinetic growth rates are similar to the MHD
growth rates while for $\beta \gg 1$ they differ significantly.  For
$\beta \gg 1$, the fastest growing mode has a growth rate $\approx
\sqrt{3} \Omega$ for a Keplerian disk, {larger} than its MHD
counterpart; there are also many modes whose growth rates are
negligible, $\lsim \beta^{-1/2} \Omega \ll \Omega$.  We provide a
detailed physical interpretation of these results and show that gas
pressure forces, rather than just magnetic forces, are central to the
behavior of the magnetorotational instability in a collisionless
plasma.  We also discuss the astrophysical implications of our
analysis.

\

\noindent {\it Subject Headings:} accretion, accretion disks;
instabilities; plasmas

\end{abstract}

\section{Introduction}

Balbus \& Hawley (1991; BH91) showed that differentially rotating
accretion disks are linearly unstable in the presence of a weak
magnetic field (see Balbus \& Hawley 1998 for a review; BH98).  This
instability, known as the ``magnetorotational instability'' (MRI), is
local and extremely powerful, with a growth rate comparable to the
rotation frequency of the disk.  MHD turbulence resulting from the MRI
is the most promising source of the efficient angular momentum
transport needed in astrophysical accretion flows (e.g., Hawley,
Gammie, \& Balbus 1995; Armitage 1998; Hawley 2000; Stone \& Pringle
2001).  The MRI may also be important for the dynamo generation of
galactic and stellar magnetic fields.

In this paper we present a linear analysis of the MRI in a
collisionless plasma using kinetic theory.  The kinetic calculation is
appropriate whenever the wavelength of the unstable modes is shorter
than the collisional mean free path. This regime is astrophysically
interesting for several reasons:

\noindent (1) In MHD, the most unstable mode of the MRI has a
wavelength $\lambda \approx v_A/\Omega$, where $v_A = B/\sqrt{4 \pi
\rho}$ is the Alfv\'en speed and $\Omega$ is the rotation frequency of
the disk.  Thus, for a very weak magnetic field, the fastest growing
mode has a very short wavelength, less than the collisional mean free
path in many cases.  A kinetic treatment is therefore required to
determine whether the MRI can amplify very weak fields (e.g., the
``first'' magnetic fields generated at high redshift by a Biermann
battery or analogous mechanism).  

\noindent (2) Radiatively inefficient accretion flows onto compact
objects provide a useful framework for interpreting observations of
low-luminosity X-ray binaries and active galactic nuclei (see, e.g.,
Ichimaru 1977; Rees et al. 1982 [the ion torus model]; Narayan \& Yi
1995 [ADAFs]; for a review see Narayan et al. 1998 or Quataert 2001).
In such models, the accreting gas is a hot two-temperature plasma in
which the proton temperature ($\sim 10^{12}$ K near a black hole) is
much larger than the electron temperature ($\sim 10^9-10^{11}$ K).  In
order to maintain such a two-temperature configuration, the accretion
flow must be effectively collisionless in the sense that the timescale
for electrons and protons to exchange energy by Coulomb collisions is
longer than the inflow time of gas in the accretion disk.  In
principle, a kinetic treatment of the accretion flow structure, rather
than a fluid treatment, is therefore necessary.  The calculations
described in this paper represent a first step towards understanding
the physics of angular momentum transport and the structure of the
accretion flow using kinetic theory.

Our analysis is restricted to wavelengths much larger than the proton
Larmor radius and frequencies below the proton cyclotron frequency.
To motivate why kinetic effects can be important even on these
``large'' scales, consider a uniform medium threaded by a weak
magnetic field ($\beta \gg 1$, where $\beta$ is the ratio of the gas
pressure to the magnetic pressure).  There are three long-wavelength
waves in such a system: (1) the sound wave, (2) the Alfv\'en wave, and
(3) the slow magnetosonic wave.  It is well known that the sound wave
and the slow wave are very different in a collisionless plasma than in
collisional plasmas described by MHD (e.g., Barnes 1966). We shall see
that the same is true for the MRI.


This paper is organized as follows.  In the next section we describe
our basic equations and assumptions (\S2).  In \S3 we discuss linear
waves in a collisionless plasma, emphasizing an important difference
between MHD and kinetic theory that is useful for understanding the
kinetic MRI results.  In \S4 we numerically solve the kinetic MRI
dispersion relation and discuss its physical interpretation. We also
show that a generalization of Balbus \& Hawley's (1992; BH92)
``spring'' model of the MRI captures the main results of the kinetic
calculation.  Finally, in \S5 we summarize our results and discuss
their astrophysical implications.

\section{Basic Equations}

In the limit that all fluctuations of interest have wavelengths much
larger than the proton Larmor radius and frequencies much less than
the proton cyclotron frequency, a collisionless plasma can be
described by the following fluid equations (e.g., Kulsrud 1983;
Snyder, Hammett, \& Dorland 1997): \beq {\partial \rho \over \partial
t} + {\bf \nabla} \cdot (\rho {\bf V}) = 0, \label{mass} \eeq \beq
\rho {\partial{\bf V} \over \partial t} + \rho ({\bf V \cdot \nabla})
{\bf V} = { ({\bf \nabla} \times {\bf B}) \times {\bf B} \over 4 \pi}
- {\bf \nabla \cdot P} + {\bf F_g}, \label{mom} \eeq \beq {\partial
{\bf B} \over \partial t} = {\bf \nabla \times (V \times B})
\label{ind}, \eeq and \beq {\bf P} = \pp {\bf I} + (\ppar - \pp) {\bf
\hat b \hat b},
\label{pressure} \eeq where $\rho$ is the mass density, ${\bf V}$ is 
the fluid velocity, $\bf B$ is the magnetic field vector, ${\bf F_g}$
is the force due to gravity, ${\bf b} = {\bf B}/|{\bf B}|$ is a unit
vector in the direction of the magnetic field, and $\bf I$ is the unit
tensor.  Equations (\ref{mass})-(\ref{ind}) are identical to the basic
equations of (collisional) MHD except that the pressure, ${\bf P}$, is
a tensor that is generally different perpendicular ($\pp$) and
parallel ($\ppar$) to the background magnetic field (e.g., the
temperature need not be isotropic in a collisionless plasma).
Formally, the pressure in equation (\ref{pressure}) should contain a
sum over all particle species in the plasma (electrons, protons, and
ions).  In what follows, however, we consider a single fluid model in
which only one contribution to the pressure response is included.  In
practice, the ions dominate the dynamics under consideration and so
the pressure can be interpreted as the ion pressure.  This is
particular true for hot accretion flows in which $T_p \gg T_e$.

In a collisionless plasma, the parallel and perpendicular pressures
satisfy separate ``equations of state'' given by (e.g., Chew,
Goldberger, \& Low 1956): \beq \rho B {d \over dt}\left({\pp \over \rho
B}\right) = -{\bf \nabla \cdot} ({\bf \hat b} \qp)- \qp {\bf \nabla
\cdot \hat b}
\label{pp} \eeq and \beq {\rho^3 \over B^2} 
{d \over dt}\left({\ppar B^2 \over \rho^3} \right) = -{\bf \nabla
\cdot} ({\bf \hat b} \qpar)- 2 \qp {\bf \nabla \cdot \hat b},
\label{ppar} \eeq where $d/dt = \partial/\partial t + {\bf V \cdot
\nabla}$ is the Lagrangian derivative, and $\qp$ and $\qpar$ represent
the flow of heat due to the motion of particles along magnetic field
lines.  Note that although there is no heat flow perpendicular to the
magnetic field due to the very small proton Larmor radius, the
perpendicular pressure/temperature can change due to heat transport
{along} the magnetic field and so $\qp \ne 0$.  If one neglects the
heat flux terms, equations (\ref{pp}) and (\ref{ppar}) reduce to
``double adiabatic theory'' (Chew et al. 1956).  Equation (\ref{pp})
then describes the invariance of the average magnetic moment of the
plasma, $\mu \propto T_\perp/B \propto \pp/(\rho B)$, where $T_\perp$
is the perpendicular temperature.  And equation (\ref{ppar}) describes
adiabatic parallel pressure changes due to the expansion or
contraction of fluid elements (Kulsrud 1983).

Equations (\ref{mass})-(\ref{ppar}) can be rigorously derived by
expanding the Vlasov equation in the long wavelength, low frequency
limit, and taking velocity moments (e.g., Kulsrud 1983).  They face,
however, the usual problem that the heat fluxes $\qp$ and $\qpar$
depend on the third moments of the particle distribution function and
so additional equations are needed to ``close'' the moment hierarchy.
In Kulsrud's kinetic MHD one avoids this closure problem by solving
the drift-kinetic equation, which is the low-frequency,
long-wavelength, limit of the Vlasov equation (see Kulsrud 1983).  By
taking moments of the resulting distribution function one calculates
$p_\perp$ and $p_\parallel$ for use in equation (\ref{mom}).  For
linear problems this approach is not too difficult and is the one
employed here (see eqs. [\ref{dperp}] and [\ref{dpar}] below).  For
nonlinear problems, however, it is much more involved.  Snyder et
al. (1997) developed fluid approximations for $\qp$ and $\qpar$ that
model kinetic effects such as Landau damping and phase mixing.  In
this approach one solves equations (\ref{pp}) and (\ref{ppar}) instead
of solving for the full distribution function.  For nonlinear
problems, this is computationally more efficient and is a possible way
of using MHD codes to extend the linear results of this paper to the
nonlinear regime.\footnote{Although we use the full drift-kinetic
equation to calculate $\pp$ and $\ppar$, we have also found that the
closures in Snyder et al. (1997) provide an excellent approximation
for the linear problems considered here.}


\subsection{Linear Perturbations}

We assume that the background (unperturbed) plasma is described by a
non-relativistic Maxwellian distribution function with equal parallel
and perpendicular pressures (temperatures).  Although the equilibrium
pressure is assumed to be isotropic, the perturbed pressure will not
be, which is a crucial difference between the kinetic and MHD
problems.  We take the plasma to be differentially rotating, but
otherwise uniform (e.g., we neglect temperature and density
gradients).  Thus the velocity is given by ${\bf V = V_0 + \delta v}$
where ${\bf V_0} = R\Omega {\bf \hat \phi}$ and $\Omega(R)$ is the
rotation rate.  We consider a weak (subthermal) magnetic field with
vertical ($B_z = B_0 \sin\theta$) and azimuthal ($B_\phi = B_0
\cos\theta$) components, where $\theta = \tan^{-1}[B_z/B_\phi]$ is the
angle between the magnetic field vector and the $\phi$ direction and
$B_0$ is the magnitude of the seed field. In a differentially rotating
plasma, a finite $B_R$ leads to a time-dependent $B_\phi$, which
greatly complicates the kinetic analysis (unlike in MHD, where a
time-dependent $B_\phi$ can be accounted for; e.g., BH91); we
therefore set $B_R = 0$.  Finally, we consider fluctuations of the
form $\exp[-i\omega t + i{\bf k \cdot x}]$, with ${\bf k} = k_R \hat R
+ k_z \hat z$, i.e., axisymmetric modes; we also restrict our analysis
to local perturbations for which $|{\bf k}| R \gg 1$.  Writing $\rho =
\rho_0 + \drho$, ${\bf B = B_0 + \delta B}$, $\pp = p_0 + \delta \pp$,
and $\ppar = p_0 + \delta \ppar$, and working in cylindrical
coordinates, the linearized versions of equations
(\ref{mass})-(\ref{ind}) become: \beq \omega \drho = \rho_0 {\bf k
\cdot \delta v} \label{linmass} \eeq \beq -i \omega \rho_0 \delta v_R
- \rho_0 2 \Omega \delta v_\phi = {-i k_R \over 4 \pi}\left(B_z \delta
B_z + B_\phi \delta B_\phi \right) + {i k_z B_z \delta B_r \over 4
\pi} - i k_R \dpp \label{rmom} \eeq \beq -i \omega \rho_0 \delta
v_\phi + \rho_0 \delta v_R {\kappa^2 \over 2 \Omega} = {i k_z B_z
\delta B_\phi \over 4 \pi} - i k_z \sin\theta \cos\theta\left[\dppar -
\dpp\right]
\label{phimom} \eeq \beq -i \omega \rho_0 \delta v_z = {-i k_z B_\phi
\delta B_\phi \over 4 \pi} - i k_z \left[\sin^2\theta \dppar + \cos^2
\theta \dpp \right]
\label{zmom} \eeq 
\beq \omega \delta B_r = - k_z B_z \delta v_R \label{rind} \eeq \beq
\omega \delta B_\phi = -k_z B_z \delta v_\phi - {i k_z B_z \over
\omega} {d \Omega \over d \ln R} \delta v_R + B_\phi {\bf k \cdot
\delta v} \label{phiind} \eeq \beq \omega \delta B_z = k_R B_z \delta
v_R, \label{zind} \eeq where $\kappa^2 = 4 \Omega^2 + d \Omega^2/d \ln
R$ is the epicyclic frequency.  Equations (\ref{linmass})-(\ref{zind})
are very similar to the analogous equations in BH91 except that we do
not impose incompressibility and the pressure response is anisotropic.
In particular, note that, even though we consider axisymmetric modes,
there is a pressure force in the $\phi$-momentum equation
(eq. [\ref{phimom}]) because the perturbed pressure is anisotropic
(i.e., ${\bf {\hat \phi} \cdot (\nabla \cdot P)} = i k_z P^{z \phi} =
i k_z (\dppar - \dpp) \sin\theta \cos\theta$).

To complete our system of equations we need expressions for $\dpp$ and
$\dppar$.  These can be obtained by taking the second moments of the
linearized and Fourier transformed drift-kinetic equation and are
given by (e.g., eqs. 23-25 of Snyder et al. 1997)

\beq {\dpp \over p_0} = {\drho \over \rho_0} + D_1 \left({\delta B
\over B_0}\right) \label{dperp} \eeq and \beq {\dppar \over p_0} =
{\drho \over \rho_0}+ D_2 \left({\drho \over \rho_0} - {\delta B \over
B_0}\right), \label{dpar} \eeq where $|{\bf B}| = B_0 + \delta B$,
$\delta B = {\bf \hat b_0 \cdot \delta B}$ is the parallel magnetic
field perturbation, and \beq D_1 = 1 - R(\xi), \ \ \ \ D_2 = \left[{1
+ 2 \xi^2 R(\xi) - R(\xi) \over R(\xi)} \right]. \label{D} \eeq Note
that the second terms on the right hand side of equations
(\ref{dperp}) and (\ref{dpar}) are the perpendicular and parallel
temperature perturbations.  In equation (\ref{D}), $R(\xi) = 1 + \xi
Z(\xi)$ is the plasma response function, \beq Z(\xi) = {1 \over
\sqrt{\pi}} \int dx {\exp[-x^2] \over x - \xi} \eeq is the plasma
dispersion function (e.g., Stix 1992), and $\xi = \omega/(\sqrt{2} c_0
|\kpar|)$, where $\kpar = {\bf \hat b \cdot k}$ is the wavevector
along the magnetic field and $c_0 = \sqrt{T/m}$ is the isothermal
sound speed of the particles (we have absorbed Boltzmann's constant
into $T$ so as to not cause confusion with the wavevector).

Because equations (\ref{dperp}) and (\ref{dpar}) are rather different
from the MHD equation of state it is worthwhile discussing their
physical interpretation.  Consider first fluctuations for which $\xi
\gg 1$, in which case $Z(\xi) \approx -\xi^{-1} - 0.5 \xi^{-3} - 0.75
\xi^{-5}$, $R(\xi) \approx -0.5 \xi^{-2} - 0.75 \xi^{-4}$, $D_1
\approx 1$, and $D_2 \approx 2$.  Equation (\ref{dperp}) thus reduces
to \beq {\dpp \over p_0} \approx {\delta \rho \over \rho_0} + {\db
\over B_0} \label{dperpadiabatic} \eeq and equation (\ref{dpar})
reduces to \beq {\dppar \over p_0} \approx 3 {\drho \over \rho_0} - 2
{\db \over B_0}. \label{dparadiabatic} \eeq These are the linearized
double adiabatic equations (eqs. [\ref{pp}] and [\ref{ppar}] with $\qp
= \qpar = 0$).  Not surprisingly, the adiabatic limit requires $\omega
\gg \kpar c_0$, i.e., that the fluctuation timescale is much less than
the time it takes particles to stream across the wavelength of the
mode.

In the opposite limit, $\xi \ll 1$, $Z(\xi) \approx i\sqrt{\pi} - 2
\xi$ and $R(\xi) \approx 1 + i\sqrt{\pi} \xi - 2 \xi^2$, so that $D_1
\approx D_2 \approx -i \sqrt{\pi} \xi$.  Equations (\ref{dperp}) and
(\ref{dpar}) thus reduce to \beq {\dpp \over p_0} \approx {\drho \over
\rho_0} - i \sqrt{\pi} \xi \left({\delta B \over B_0}\right)
\label{dperpsimp} \eeq and \beq {\dppar \over p_0} \approx {\drho
\over \rho_0} - i \sqrt{\pi} \xi \left({\drho \over \rho_0} - {\delta
B \over B_0}\right). \label{dparsimp} \eeq These correspond to nearly
isothermal fluctuations: the temperature perturbation is smaller than
its ``natural'' value by a factor $\sim \xi \ll 1$.  This is because
$\omega \ll \kpar c_0$, i.e., particles stream across a wavelength on
a timescale much less than the fluctuation timescale and efficiently wipe
out temperature gradients.  Equations (\ref{dperpsimp}) and
(\ref{dparsimp}) are the appropriate limit for the MRI.  This is
because the MRI has $|\omega| \lsim k_z v_A$ and $\beta \gsim 1$, and
so $\xi \lsim \beta^{-1/2} \lsim 1$.

\subsection{The Dispersion Relation}

To obtain the dispersion relation, we eliminate all non-velocity
variables from the momentum equations.  We first calculate $\delta B =
\cos\theta \delta B_\phi + \sin\theta \delta B_z$ using $\delta
B_\phi$ from equation (\ref{phiind}) and $\delta B_z$ from equation
(\ref{zind}): \beq {\delta B \over B_0} = \cos^2\theta {{\bf k \cdot
\delta v} \over \omega} - \sin\theta \cos\theta {k_z \delta v_\phi
\over \omega} - i {d \Omega \over d \ln R} \sin\theta \cos\theta {k_z
\delta v_R \over \omega^2} + \sin^2\theta {k_R \delta v_R \over
\omega}.
\label{bpar} \eeq  Substituting equations (\ref{linmass}) and 
(\ref{bpar}) into equations (\ref{dperp}) and (\ref{dpar}) then yields
$\dpp$ and $\dppar$ as functions of $\bf \delta v$, which we
substitute into the perturbed momentum equations
(eqs. [\ref{rmom}]-[\ref{zmom}]).  Using the perturbed induction
equations (eqs. [\ref{rind}]-[\ref{zind}]) we can also eliminate $\bf
\delta B$ from the momentum equations in favor of $\bf \delta v$.
This yields the following versions of the momentum equations:
\begin{eqnarray} &0& =  \ \delta v_R \Huge[\omega^2 - k^2 v_{Az}^2 -
k^2_R(v^2_{A\phi} + c_0^2) + i k_R k_z {v_{Az} v_{A \phi} \over
\omega} {d\Omega \over d \ln R} + ik_R k_z c^2_0 {D_1 \over \omega}
\sin\theta \cos \theta {d \Omega \over d \ln R} \nonumber \\ &-& k_R^2
c^2_0 D_1\Huge] \ + \ \delta v_\phi \left[k_R k_z v_{Az} v_{A \phi} -
2 i \Omega \omega + k_R k_z c^2_0 D_1 \sin\theta \cos\theta\right]
\nonumber \\ &-& \delta v_z \left[ k_R k_z (c^2_0 + v_{A\phi}^2) + k_R
k_z c^2_0 \cos^2\theta D_1 \right], \label{rfinal}
\end{eqnarray}
\begin{eqnarray}
&0& = \ \delta v_R \Huge[{i \kappa^2 \omega \over 2 \Omega} - i {k_z^2
v_{Az}^2 \over \omega} {d \Omega \over d \ln R} + v_{A\phi} v_{Az} k_R
k_z + k_z k_R c^2_0 D_1 \sin\theta \cos\theta \nonumber \\ &-& i
{c^2_0 k^2_z \over \omega}(D_2 + D_1) \sin^2\theta \cos^2\theta {d
\Omega \over d \ln R}\Huge] + \delta v_\phi\left[\omega^2 - k^2_z
v^2_{Az} - k_z^2 c_0^2 \sin^2\theta \cos^2\theta (D_2 + D_1)\right]
\nonumber \\ &+& \delta v_z \left[k^2_z v_{A\phi} v_{Az} - k_z^2 c^2_0
D_2 \sin^3\theta \cos \theta + k_z^2 c^2_0 D_1 \sin\theta \cos^3\theta
\right] \label{phifinal}
\end{eqnarray}
\begin{eqnarray}
&0& = \ \delta v_R \Huge[ -k_R k_z(v^2_{A\phi} + c^2_0) + i k^2_z
{v_{A\phi} v_{Az} \over \omega} {d \Omega \over d \ln R} - k_z k_R
c^2_0 D_1 \cos^2\theta + i {k_z^2 c_0^2 \over \omega} D_1 \cos^3\theta
\sin\theta {d \Omega \over d \ln R} \nonumber \\ &-& i {k^2_z c^2_0
\over \omega} D_2 \sin^3\theta \cos\theta {d \Omega \over d \ln R}
\Huge] + \delta v_\phi \left[k^2_z v_{A\phi} v_{Az} - k_z^2 c^2_0 D_2
\sin^3\theta \cos\theta + k_z^2c_0^2 D_1 \cos^3\theta \sin\theta
\right] \nonumber \\ &+& \delta v_z\left[\omega^2 - k^2_z(v_{A\phi}^2
+ c_0^2) - k^2_z c_0^2 D_2 \sin^4\theta - k^2_z c^2_0 D_1 \cos^4\theta
\right] \label{zfinal}, \end{eqnarray} where \beq v_{Az}^2 \equiv
{B^2_z \over 4 \pi \rho_0} \ \ \ {\rm and} \ \ \ v_{A\phi}^2 \equiv
{B^2_\phi \over 4 \pi \rho_0} \eeq are the Alfv\'en speeds associated
with the vertical and azimuthal fields, respectively.  Equations
(\ref{rfinal})-(\ref{zfinal}) define a matrix equation of the form $A
{\bf \delta v} = 0$.  Setting det(A) = 0 gives the dispersion
relation.  We have not found it particularly illuminating to write out
the entire dispersion relation, nor have we made much progress solving
it analytically, so instead we proceed to discuss its numerical
solution.  We will also present a simple model problem that captures
the essential physics of the kinetic MRI.  

The MHD dispersion relation for the MRI, including the effects of
compressibility, can be obtained from equations
(\ref{rfinal})-(\ref{zfinal}) by setting $D_1 = D_2 = 0$.  Equations
(\ref{dperp}) and (\ref{dpar}) show that, for $D_1 = D_2 = 0$, $\dpp =
\dppar = \drho c_0^2$, i.e., the perturbations are isothermal and the
perturbed pressure is isotropic.  Our basic linear perturbation
equations (eqs. [\ref{linmass}]-[\ref{zind}]) reduce to their MHD
analogues in this limit. In particular, note that in MHD the MRI is
independent of whether the perturbations are adiabatic or isothermal;
this is because it is an incompressible instability so the precise
form of the sound speed is irrelevant for $\beta \gg 1$ (e.g., BH91).
Thus the key simplification to the kinetic equations obtained by
setting $D_1 = D_2 = 0$ is that the perturbed pressure becomes
isotropic, as it is in MHD.


\section{Linear Waves in Double Adiabatic Theory}
Before considering the full kinetic MRI problem, it is instructive to
consider the simpler problem of linear waves in a uniform medium.  In
particular, we show that the slow magnetosonic wave is very different
in kinetic theory than in MHD.  This is important for understanding
the kinetic MRI because the slow wave, along with the Alfv\'en wave,
is central to the dynamics of the MRI.  We use double adiabatic theory
throughout this section. Although double adiabatic theory does not
include collisionless damping, which is quite strong for the slow mode
and would alter some of the results in this section, it does show the
significant differences introduced by the anisotropic pressure in a
collisionless plasma.  Since our interpretation of the kinetic MRI in
\S4 focuses on the importance of this anisotropic pressure, it is
useful to see its implications first in a simpler problem.  In \S4 and
the Appendix we show that the qualitative conclusions drawn in this
section carry over to the full kinetic analysis.

Double adiabatic theory in a uniform medium corresponds to setting
$\Omega = \kappa = 0 $ and $\xi \gg 1$ in equations
(\ref{rfinal})-(\ref{zfinal}), in which case $D_1 = 1$ and $D_2 = 2$.
Without loss of generality we can take $B_\phi = 0$ so that
$\cos\theta = 0$ and $\sin\theta = 1$.  To make contact with standard
notation, we also write $k_r = k_\perp$, $k_z = k_\parallel$, and $v_{Az}
= v_{A}$.  The dispersion relation is then given by \beq
\left[\omega^2 - \kpar^2 v_A^2\right]\left[(\omega^2 - k^2 v_A^2 - 2
\kp^2 c_0^2)(\omega^2 - 3 \kpar^2 c_0^2) - \kp^2\kpar^2c^4_0\right]
\equiv D_A D_{MS} = 0. \label{dadisp} \eeq The analogous MHD
dispersion relation is \beq \left[\omega^2 - \kpar^2
v_A^2\right]\left[(\omega^2 - k^2 v_A^2 - \kp^2 v_s^2)(\omega^2 -
\kpar^2 v_s^2) - \kp^2\kpar^2v^4_s\right] = 0,
\label{mhddisp} \eeq where $v_s^2 = \gamma c_0^2$ is the adiabatic sound
speed and $\gamma = 5/3$ is the adiabatic index.

Equation (\ref{dadisp}) shows that, as in MHD, the double adiabatic
dispersion relation factors into two parts: an Alfv\'en wave branch
($D_A = 0$) and a magnetosonic branch ($D_{MS} = 0$).  The Alfv\'en
wave in double adiabatic theory is identical to that in MHD, while the
magnetosonic waves are different -- this is because the ``adiabatic
index'' in a collisionless plasma is different for motions
perpendicular and parallel to the magnetic field.  Motion along the
field is one-dimensional and corresponds to $\gamma = 3$ (hence the
$3\kpar^2$ term in eq. [\ref{dadisp}]) while motion perpendicular to
the field is two-dimensional and corresponds to $\gamma = 2$ (hence
the $2\kp^2$ term in eq. [\ref{dadisp}]).  By contrast, in MHD, the
pressure is isotropic and $\gamma = 5/3$.

Figure 1 shows a plot of the dispersion relation of the fast and slow
magnetosonic waves in MHD (dotted lines) and in double adiabatic
theory (solid lines), taking $\beta = 100$.  The fast wave, which is
essentially a sound wave, is qualitatively similar in the two cases
(the quantitative differences are due to the different $\gamma$'s).
The slow wave, however, is quite different. In MHD, the dispersion
relation of the slow wave is degenerate with that of the Alfv\'en wave
for $\beta \gg 1$, namely $\omega = \kpar v_A$.  Except for $\kp = 0$,
this is not true in double adiabatic theory.  The frequency of the
slow wave depends on the sound speed; in fact, for $\kp \ne 0$, the
primary restoring force for the slow wave in double adiabatic theory
is gas pressure, not magnetic forces.

This result can be understood as follows.  In MHD, the properties of
the $\beta \gg 1$ slow wave can be calculated by explicitly imposing
incompressibility, ${\bf \nabla \cdot \delta v} \propto \delta \rho
\approx 0$.  This additional constraint (incompressibility) replaces
the equation of state to determine the pressure (the Boussinesq
approximation).  In a collisionless plasma, this cannot happen because
the pressure response is different parallel and perpendicular to the
magnetic field, i.e., there are {\it two} equations of state (one for
$\pp$ and one for $\ppar$).  Both equations of state cannot be
replaced by the single requirement that the fluctuations be
incompressible.  More physically, a $\kp \sim k_\parallel$ slow wave
in MHD has $\delta p \sim \delta B^2/8\pi \sim B_0 \ \delta B_\perp/ 4
\pi$, i.e., the gas pressure, magnetic pressure, and magnetic tension
forces are all comparable.  Equivalently, $\delta p/p_0 \sim
\beta^{-1} \delta B / B_0 \ll \delta B/B_0$ (for $\beta \gg 1$).  In
double adiabatic theory, however, a parallel magnetic field
perturbation $\delta B/B_0$ induces a pressure perturbation $\delta
p_{\perp,\parallel}/p_0$ of comparable magnitude (see
eqs. [\ref{dperpadiabatic}] and [\ref{dparadiabatic}]).  This means
that the pressure forces are much larger than the magnetic forces
($\delta p \sim \beta \ \delta B^2/8 \pi \gg \delta B^2/8 \pi$) and
dominate the dynamics of the wave.

The exception to these arguments is if the pressure perturbation
vanishes, i.e., $\dpp = \dppar = 0$.  Alfve\'n waves and the $\kp
\rightarrow 0$ limit of the slow magnetosonic wave are the only waves
in MHD that have $\delta p = 0$ (they also have $\delta B = 0$).  As a
result, these pressure-free waves are the only incompressible
fluctuations in double adiabatic theory.  For all other waves, and in
particular for slow waves with $\kp \ne 0$, pressure is the dominant
restoring force in a $\beta \gg 1$ plasma and so the frequencies
depend strongly on the sound speed (Fig. 1).

The results in this section are relevant to the MRI because the MRI is
an incompressible instability with $|\omega| \ll k c_0$.  Although
pressure forces generally lead to a small modification of the MRI in
MHD, they will be substantially more important in the kinetic analysis
(just as for the slow magnetosonic wave in this section).
\section{The Kinetic MRI}

As noted in \S2.3, the general kinetic MRI dispersion relation appears
to be analytically intractable.  In this section we present its
numerical solution and physical interpretation.  As a check on our
numerical calculations we have confirmed that our results reproduce
the kinetic dispersion relation for the Alfv\'en wave and the slow and
fast magnetosonic waves when $\Omega = 0$ (including the collisionless
damping rates).\footnote{We compared our results to the linear kinetic
code described in Quataert (1998) and to the analytic results in
Barnes (1966) and Foote \& Kulsrud (1979).}  We also reproduce the MRI
in MHD when the kinetic terms are dropped (this requires setting $D_1
= D_2 = 0$ in eqs. [\ref{rfinal}]-[\ref{zfinal}]).

Figures 2-4 show the results of numerically solving the kinetic MRI
dispersion relation, assuming a Keplerian disk for which $\Omega
\propto R^{-3/2}$ and $\kappa = \Omega$.  The figures show the kinetic
growth rate of the MRI for different values of $\beta_z \equiv 8 \pi
p_0/B^2_z$, for different magnetic field geometries (defined by
$B_\phi/B_z$), and for different wavevectors ($k_R$ and $k_z$).  The
corresponding MHD results are shown for comparison by the dotted
lines.  It is important to note that in MHD the MRI growth rate is
essentially independent of $\beta$ and $B_\phi/B_z$; by contrast,
Figures 2-4 show that the kinetic results depend sensitively on both
of these parameters.

Figures 2-4 show that, although the growth rates can be very
different, the region of instability in wavevector space is the same
in kinetic theory and MHD.  To understand this result, it is
sufficient to consider the $\omega \rightarrow 0$ limit of the kinetic
equations, since this determines the transition between stable and
unstable modes.  Setting $\omega = 0$ implies that $\xi \equiv
\omega/(\sqrt{2} \kpar c_0)= 0$ as well.  From equations
(\ref{dperp})-(\ref{D}), it then follows that $\dpp/p_0 = \dppar/p_0 =
\drho/\rho_0$.  Physically, as $\xi \rightarrow 0$, there is more and
more time for particles moving along magnetic field lines to
efficiently transport heat.  This leads to nearly isothermal
fluctuations in which the pressure perturbation is isotropic and is
set only by the density perturbation.  As discussed in \S2.3, the
kinetic equations reduce to the MHD equations in this limit.  This is
an important result because it shows that the MHD instability
criterion, namely $d\Omega^2/dR < 0$, applies to the kinetic problem
as well.  Moreover, the set of unstable modes is the same in MHD and
kinetic theory, as is seen explicitly in Figures 2-4.

Perhaps the three most striking results of the kinetic calculation
shown in Figures 2-4 are: (1) The kinetic growth rates depend
sensitively on $\beta$.  For $\beta \gg 1$ they differ significantly
from the MHD growth rates while for $\beta \sim 1$ they are similar
(see Fig. 3).  (2) For $B_\phi = 0$, or for sufficiently large $k_R$,
the kinetic growth rates are smaller than their MHD counterparts,
particularly at large $\beta$ (e.g., Fig. 3a).  (3) For $B_\phi \ne
0$, the kinetic growth rates can be {\it larger} than their MHD
counterparts (e.g., Fig. 2 and Fig. 4b).  Moreover, for $\beta \gg 1$,
the fastest growing mode is at $k_z v_{Az} \ll \Omega$, where there is
negligible growth in MHD (Fig. 4b).

To understand the kinetic MRI results, we have found it useful to
consider the equations that describe the displacement of a fluid
element from its equilibrium circular orbit.  BH92 and BH98 showed
that, for the special case of a vertical magnetic field and vertical
wavevector, the radial and azimuthal components of the MHD momentum
equation can be written in terms of the radial and azimuthal fluid
displacements, $\xi_R$ and $\xi_\phi$, as \beq {\partial^2 \xi_R \over
\partial t^2} - 2 \Omega {\partial \xi_\phi \over \partial t } =
-\left({d \Omega^2 \over d \ln R} + (k_zv_{Az})^2\right)\xi_R,
\label{radial} \eeq \beq {\partial^2 \xi_\phi \over \partial t^2} + 2 
\Omega {\partial \xi_R \over \partial t } = -(k_z v_{Az})^2
\xi_\phi. \label{azimuthal} \eeq As discussed by BH92 and BH98,
equations (\ref{radial}) and (\ref{azimuthal}) are identical to the
equations describing two orbiting point masses connected by a spring
of spring constant $k_z^2 v_{Az}^2$ (in MHD, magnetic tension plays
the role of the spring).  This suggests the following physical
interpretation of the MRI in MHD (BH92).  For a rotation profile with
$d \Omega^2/d R < 0$ (unstable to the MRI), a fluid element at radius $R
- \delta R$ is rotating slightly faster than a fluid element at radius
$R$.  The ``spring'' pulls backwards on this inner fluid element,
removing its angular momentum and forcing it to move to a yet smaller
radius.  Similarly, a fluid element at radius $R + \delta R$ is
rotating slightly slower than a fluid element at radius $R$ and so the
``spring'' pulls forward on this fluid element, giving it angular
momentum and forcing it to move to a yet larger radius.  This simple
physical picture captures the essence of the MRI in MHD.

A useful toy model that provides additional insight into the physics
of the MRI, both in MHD and kinetic theory, is given by the following
equations for the fluid displacement \beq {\partial^2 \xi_R \over
\partial t^2} - 2 \Omega {\partial \xi_\phi \over \partial t } =
-\left({d \Omega^2 \over d \ln R} + K_R \right)\xi_R, \label{radtoy}
\eeq \beq {\partial^2 \xi_\phi \over \partial t^2} + 2 \Omega
{\partial \xi_R \over \partial t } = -K_\phi \xi_\phi. \label{aztoy}
\eeq Equations (\ref{radtoy}) and (\ref{aztoy}) describe the
displacement of rotating fluid elements coupled by an anisotropic
``spring,'' for which the spring constant is different in the
azimuthal ($K_\phi$) and radial ($K_R$) directions (this is clearly no
longer a real spring!).  The unstable root in the dispersion relation
associated with equations (\ref{radtoy}) and (\ref{aztoy}) is given by
\beq \omega^2 = {\kappa^2 + K_R + K_\phi \over 2} - {1 \over 2} \left[
\left(K_\phi + K_R + \kappa^2\right)^2 - 4 K_\phi\left(K_R + {d
\Omega^2 \over d \ln R} \right)\right]^{1/2}. \label{toydisp} \eeq For
$K_R = K_\phi = (k_z v_{Az})^2$, equation (\ref{toydisp}) gives the
$k_R = 0$ dispersion relation of the MRI in MHD (this is shown by the
dotted line in Fig. 4b).  It is also straightforward to show that, for
$K_R > K_\phi$, the growth rates in equation (\ref{toydisp}) are
smaller than the MHD growth rates (i.e., those with $K_R = K_\phi$)
and for $K_\phi > K_R$ the growth rates in equation (\ref{toydisp})
are larger than the MHD growth rates.  For example, for $K_\phi = 0$,
equation (\ref{toydisp}) gives $\omega = 0$ for any $K_R$ and so there
is no instability.  On the other hand, the $K_\phi \gg \Omega \gg K_R$
solution of equation (\ref{toydisp}) is $|\omega| \approx \sqrt{|d
\Omega^2/d \ln R|}$.  For a Keplerian disk this gives $|\omega| =
\sqrt{3} \Omega$, which is larger than the growth rate of the fastest
growing mode in MHD ($|\omega| = 3 \Omega/4$).

These results can be understood physically by noting that it is
ultimately the presence of an azimuthal restoring force, rather than a
radial restoring force, that is {\it destabilizing} in the MRI.  This
is because it is the azimuthal force that removes angular momentum
from an inwardly displaced fluid element and adds it to an outwardly
displaced fluid element.  By contrast, the radial force is stabilizing
because it attempts to ``pin'' the fluid element to its equilibrium
position.  Thus $K_\phi > K_R$ leads to faster growth because it
enhances the destabilizing azimuthal force relative to the stabilizing
radial force (and vice-versa for $K_R > K_\phi$).  For the remainder
of this section we explain how thermal pressure in a collisionless
plasma plays the role of the anisotropic ``spring'' in the above toy
model.  This will account for the behavior of the kinetic MRI seen in
Figures 2-4.

Because they are restricted to $k_R = 0$ and $B_\phi = 0$, equations
(\ref{radial}) and (\ref{azimuthal}) do not include the effects of gas
pressure or magnetic pressure (both of which vanish in this special
case).  To understand the kinetic MRI we need to include these
restoring forces using the radial and azimuthal momentum equations
(eqs. [\ref{rmom}] and [\ref{phimom}], respectively).  This yields the
following equations for the fluid displacement\footnote{Strictly
speaking, equations (\ref{azimuthal}) and (\ref{azi}) should have an
additional term on the right hand side given by $k_z v_{Az} v_{A\phi}
({\bf k \cdot \xi})$.  For $\beta \gg 1$, this term is negligible
because the MRI is nearly incompressible and so we do not consider it
further.} \beq {\partial^2 \xi_R \over \partial t^2} - 2 \Omega
{\partial \xi_\phi \over \partial t } = -\left({d \Omega^2 \over d \ln
R} + (k_zv_{Az})^2\right)\xi_R - ik_R\left({\delta B^2 \over 8 \pi
\rho_0} + {\dpp \over \rho_0}\right),
\label{rad} \eeq \beq {\partial^2 \xi_\phi \over \partial t^2} + 2 
\Omega {\partial \xi_R \over \partial t } = -(k_z v_{Az})^2 \xi_\phi -
i k_z\left({\dppar - \dpp \over \rho_0}\right) \sin\theta
\cos\theta. \label{azi} \eeq In equations (\ref{rad}) and (\ref{azi})
we have simply rewritten the pressure gradients from equations
(\ref{rmom}) and (\ref{phimom}); in the Appendix we calculate these
explicitly in terms of the fluid displacement.  It is worth noting
again that there is a pressure force in the $\phi$-momentum equation
(eq. [\ref{azi}]) because the perturbed pressure is anisotropic.  In
MHD, $\dppar = \dpp$ and so this term vanishes.

Following the arguments in \S3 and Figure 1 we expect that the
pressure gradients in equations (\ref{rad}) and (\ref{azi}) will be
much more important in kinetic theory than in MHD.  In the Appendix we
calculate the magnitude of these pressure forces and confirm this
hypothesis. We use these results below to present a physical
interpretation of the kinetic MRI, focusing on two important special
cases: (1) $B_\phi = 0$, $k_R \ne 0$, for which the kinetic growth
rates are smaller than their MHD counterparts (e.g., Figs. 3a \& 4a),
and (2) $k_R = 0$, $B_\phi \ne 0$, for which the kinetic growth rates
are larger than the MHD growth rates (e.g., Fig. 4b).

Consider first the special case of $B_\phi = 0$ and $k_R \ne 0$ (e.g.,
Fig. 3a \& 4a).  In this case a displaced fluid element feels a
restoring force in the radial direction due to gas and magnetic
pressure; there is, however, no analogous pressure gradient in the
$\phi$ direction (only magnetic tension).  This corresponds to $K_R >
K_\phi$ in the toy model of equation (\ref{toydisp}); the growth rates
should therefore be suppressed with respect to the $k_R = 0$ growth
rates.  The presence of a stabilizing radial pressure gradient
provides a physical explanation for why the MHD growth rates decrease
with increasing $k_R$ (see, e.g., the dotted line in Fig. 3a).
Moreover, in the Appendix we show that the pressure gradient in
kinetic theory is larger than in MHD by a factor of $\sim
\beta^{1/2}$.  The kinetic growth rates should therefore be even
smaller than the MHD growth rates, with stronger suppression at larger
$\beta$.  This is precisely what is seen in the kinetic calculation;
e.g., Figures 3a and 4a show the $B_\phi = 0$ growth rate for
different $\beta$.

Consider now the special case of $k_R = 0$, but $B_\phi \ne 0$ (e.g.,
Fig. 4b).  In this case the radial pressure force vanishes, but there
is an azimuthal pressure force due to the anisotropic pressure.  As
suggested by the toy model in equation (\ref{toydisp}) this azimuthal
pressure force, which is not present in MHD, is {\it destabilizing}
because it removes angular momentum from an inwardly displaced fluid
element and adds it to an outwardly displaced fluid element (just as
the azimuthal component of magnetic tension does).  Moreover, for
$B_\phi \sim B_z$ the destabilizing pressure force is larger than the
destabilizing magnetic tension force by a factor of $\sim \beta^{1/2}$
(see the Appendix).  This explains why the $k_R = 0$, $B_\phi \ne 0$
growth rates are larger than their MHD counterparts, and why the
growth rates increase with increasing $\beta$ (see, e.g., Fig. 4b).
It also explains why the growth can be rapid even at $k_z v_{Az} \ll
\Omega$, when magnetic tension (which drives the MRI in MHD) is very
weak.  In fact, for $\beta \gg 1$ and $k_z v_{Az} \ll \Omega$, the
forces in the kinetic MRI are arranged as follows: azimuthal pressure
$\gg$ Coriolis $\gg$ magnetic tension.  We therefore expect the growth
rates to approach the $K_\phi \gg \Omega \gg K_R$ limit of equation
(\ref{toydisp}), namely $|\omega| \approx \sqrt{3} \Omega$.  As shown
in Figure 4b, the fastest growing modes do approach this maximal
growth rate.

Although the above interpretation focuses on two special cases, the
results in Figures 2-4 can be readily understood as a competition
between the stabilizing radial pressure force and the destabilizing
azimuthal pressure force.  The importance of {gas pressure}, rather
than {magnetic} forces, also explains why the kinetic results depend
sensitively on $\beta$.  



\section{Summary and Discussion}

In this paper we have presented a linear axisymmetric calculation of
the magnetorotational instability (MRI) in a collisionless plasma.
Our analysis is restricted to wavelengths much larger than the proton
Larmor radius, frequencies below the proton cyclotron frequency, and
``seed'' magnetic fields with no radial component ($B_R = 0$).  The
MRI is believed to give rise to MHD turbulence and efficient angular
momentum transport in astrophysical accretion flows, and may also be
important for the dynamo generation of galactic and stellar magnetic
fields (e.g., BH98).  Our kinetic calculation, rather than an MHD
calculation, is appropriate whenever the collisional mean free path of
the protons exceeds the wavelength of the MRI.

The instability criterion for the kinetic MRI is the same as in MHD,
namely that the angular velocity decrease outwards.  The set of
unstable modes is also the same in kinetic theory and MHD.  However,
nearly every mode has a linear kinetic growth rate that differs from
its MHD counterpart.  For example, the fastest growing mode in kinetic
theory has a growth rate $\approx \sqrt{3} \Omega$ for a Keplerian
disk, which is larger than its MHD counterpart by a factor of $4
\sqrt{3}/3 \approx 2.3$.\footnote{This rapid growth is obtained only
for $k_R v_{Az} \ll \Omega$, $k_z v_{Az} \ll \Omega$, $\beta \gg 1$,
and $B_\phi \gsim B_z$ (see Figs. 3 \& 4).}  More generally, the
kinetic growth rates can be either larger or smaller than the MHD
growth rates, depending on the orientation of the magnetic field and
the wavevector of the mode (Fig. 2).  The kinetic growth rates also
depend explicitly on $\beta$, i.e., on the ratio of the gas pressure
to the magnetic pressure.  For $\beta \gg 1$ the kinetic results
differ significantly from the MHD results while for $\beta \sim 1$
they are similar (see Fig. 3).

We have argued that the kinetic MRI can be understood by considering
the force due to pressure gradients in a high $\beta$ collisionless
plasma.  In MHD, pressure leads to a relatively minor modification of
the MRI.  In kinetic theory, however, the pressure forces are $\sim
\beta^{1/2}$ times larger than in MHD and are therefore dynamically
much more important (see \S3 and the Appendix).  Moreover, in kinetic
theory there is an azimuthal pressure force even for axisymmetric
modes (so long as $B_\phi \ne 0$; see eqs. [\ref{zmom}] and
[\ref{azi}]).  This is because the pressure response is anisotropic in
a collisionless plasma: it is different along and perpendicular to the
local magnetic field.  This azimuthal pressure force, which is not
present in MHD, is {\it destabilizing} because it removes angular
momentum from an inwardly displaced fluid element and adds it to an
outwardly displaced fluid element (just as the azimuthal component of
magnetic tension does in MHD).  The destabilizing pressure force
explains why the kinetic growth rates of the MRI can be larger than
the MHD growth rates (e.g., Fig. 4b).

The importance of gas pressure shows that the character of the MRI is
somewhat different in a collisionless plasma than in a collisional
plasma described by MHD.  The crucial function of the magnetic field
is to enforce an anisotropic pressure response, rather than to
directly destabilize the plasma via magnetic tension.  The importance
of pressure gradients also explains why the kinetic results depend
sensitively on $\beta$.  For $\beta \sim 1$ pressure forces are
comparable to magnetic forces, and the kinetic growth rates are not
that different from the MHD growth rates, while for $\beta \gg 1$
pressure forces dominate over magnetic forces and the kinetic results
differ substantially from the MHD results (e.g., Fig. 3).

BH92 showed that the MRI in MHD could be understood using a simple
model in which magnetic tension acts like a spring coupling different
fluid elements in the plasma.  We have presented a generalization of
BH92's ``spring'' model that captures many of the results of the full
kinetic MRI calculation (see eqs. [\ref{radtoy}]-[\ref{toydisp}]).  In
this model the radial and azimuthal ``spring constants'' are
different; physically, this corresponds to the anisotropic pressure
response in a collisionless plasma.

To conclude, we briefly discuss the astrophysical implications of our
results, focusing on the two applications mentioned in the
introduction: (1) the amplification of weak fields generated by a
Biermann battery or analogous mechanism, and (2) hot two-temperature
accretion flows onto compact objects.

\noindent (1) For a very weak magnetic field MHD predicts that the
fastest growing mode of the MRI has a very small wavelength $\approx
v_A/\Omega \propto B$.  This will be less than the collisional mean
free path in many cases.  Our kinetic analysis shows that there is
rapid growth of the MRI even in this limit.  This is encouraging for
the hypothesis that the MRI contributes to the dynamo amplification of
very weak magnetic fields, e.g. the generation of galactic fields from
a cosmological seed field.  To further assess this question, however,
it is necessary to extend our analysis to include finite Larmor radius
effects.  In particular, the ``battery'' generation of magnetic fields
is limited by self-induction to field strengths such that the proton
Larmor radius is comparable to the size of the system (e.g., Balbus
1993).  Finite Larmor radius effects will always be important on these
scales, particularly since the wavelengths of unstable MRI modes are
then much less than the proton Larmor radius.

\noindent (2) In radiatively inefficient accretion flows onto compact
objects, which have been applied extensively to low-luminosity
accreting sources (e.g., Narayan et al. 1998), the inflowing gas is a
hot two-temperature plasma in which the proton temperature is much
larger than the electron temperature.  In order to maintain $T_p \gg
T_e$, the timescale for electrons and protons to exchange {energy} by
Coulomb collisions must be longer than the inflow time of the gas.
This requires a sufficiently small accretion rate, $\dot M \lsim
\alpha^2 \dot M_{EDD}$ (e.g., Rees et al. 1982), where $\dot M_{EDD}$
is the Eddington accretion rate and $\alpha$ is the dimensionless
Shakura-Sunyaev viscosity parameter.  Since the timescale for
proton-electron collisions to modify the proton distribution function
is comparable to the proton-electron energy exchange timescale, the
proton dynamics is effectively collisionless for {\it any}
two-temperature radiatively inefficient accretion
flow;\footnote{Proton-electron collisions are more important than
proton-proton collisions because $T_p \gg T_e$.}  the kinetic
calculation presented in this paper is therefore appropriate for
describing angular momentum transport by the MRI in such
models.\footnote{ By contrast, geometrically thin accretion disks
(e.g., Shakura \& Sunyaev 1973) are much cooler and denser; MHD
accurately describes the dynamics of thin disks so long as the gas is
sufficiently ionized.}

It is, however, difficult to apply our linear calculations to the
nonlinear saturated state expected in the accretion flow.  Nonetheless
it is worth noting that there are rapidly growing modes in a
collisionless plasma even for $\beta \gg 1$ so weak fields can be
efficiently amplified.  Moreover, MHD simulations find saturation at
$\beta \sim 1-100$ with a predominantly toroidal field (e.g., BH98;
Stone \& Pringle 2001).  For this magnetic field configuration, the
linear kinetic growth rates of the MRI are not that different from
their MHD counterparts (if anything, they may be somewhat larger;
e.g., Figs. 3b \& 4b).  While this suggests that the saturated
turbulence may be qualitatively similar in kinetic theory and MHD,
there will undoubtedly be quantitative differences.  In addition, the
fact that the fastest growing modes occur at somewhat different
wavenumbers could change the nonlinear results.  Perhaps more
importantly, collisionless damping of the sound wave and the slow
magnetosonic wave is very strong and operates on all scales in a
collisionless plasma, while strong damping in MHD is restricted to
very small scales.  This may alter the nonlinear behavior of the MRI.
Numerical simulations that address these issues would be extremely
interesting.

Our results may also have implications for understanding particle
heating in radiatively inefficient accretion flows.  The radiative
efficiency of such models is set by the amount of electron heating in
the plasma.  This depends on how the energy in MHD turbulence is
dissipated (e.g., via a turbulent cascade, reconnection, etc.).  The
prominent role of pressure fluctuations in the kinetic MRI suggests
that the resulting turbulence may couple better to slow waves (which
have a pressure perturbation) than Alfv\'en waves (which do not).
Slow waves primarily heat the protons in the collisionless plasmas of
interest (e.g., Quataert 1998; Blackman 1999) while an Alfv\`enic
cascade may lead to significant electron heating if $\beta \lsim 10$
(e.g., Gruzinov 1998; Quataert \& Gruzinov 1999).  Kinetic simulations
of the MRI should be able to assess the relative importance of slow
wave and Alfv\'en wave excitation.

\acknowledgements We thank Steve Balbus, Steve Cowley, Barrett Rogers,
Alex Schekochihin, and Anatoly Spitkovsky for useful discussions.  GH
was supported in part by the U.S. Department of Energy under Contract
DE-AC02-76CH03073.  EQ was supported in part by NASA grant NAG5-12043.

\newpage

\begin{appendix}

\section{Calculation of the Pressure Forces}

In this Appendix we calculate the radial and azimuthal pressure forces
in equations (\ref{rad}) and (\ref{azi}) in terms of the fluid
displacements $\xi_R$ and $\xi_\phi$.  These are used in our
interpretation of the kinetic MRI results in \S4.  We restrict our
analysis to the two important limits highlighted in \S4: (1) $B_\phi =
0; k_R \ne 0$ and (2) $B_\phi \ne 0; k_R = 0$. 

\subsection{$B_\phi = 0; k_R \ne 0$} 
In this case there is a radial pressure force given by
(eq. [\ref{rad}])\beq F_R \equiv -i k_R\left({\delta B^2 \over
8\pi\rho_0} + {\dpp \over \rho_0} \right) = -i k_R \left[{\drho \over
\rho_0}c^2_0 + {\delta B \over B_0}\left(v_{Az}^2 + D_1
c^2_0\right)\right],
\label{frapp} \eeq where we have used $\dpp/p_0 = \drho/\rho_0 + D_1 
\delta B/B_0$ from equation (\ref{dperp}) in the second equality.  We
now rewrite all of the terms in $F_R$ in terms of $\xi_R$, the radial
displacement. For $B_\phi = 0$, $\delta B = \delta B_z$.  The radial
component of the induction equation (eq. [\ref{rind}]) thus yields
\beq {\delta B \over B_0} = {k_R \delta v_R \over \omega} = -i k_R
\xi_R. \label{dbapp} \eeq 

To calculate $\drho/\rho_0 = (k_z \delta v_z + k_R \delta v_R)/\omega$
in terms of $\xi_R$ alone we need to find $\delta v_z$ as a function
of $\delta v_R$.  To do this note that the z-component of the momentum
equation (eq. [\ref{zmom}]) implies \beq k_z \delta v_z = {k_z^2
\dppar \over \omega \rho_0}. \label{vz} \eeq Since $\drho/\rho_0 -
\delta B/B_0 = k_z \delta v_z/\omega$, equation (\ref{dpar}) gives
$\dppar$ as a function of both $\delta v_z$ and $\delta v_R$.
Substituting this into equation (\ref{vz}) we solve for $\delta v_z$
in terms of $\delta v_R$ and thus find \beq {\drho \over \rho_0} = - i
k_R \xi_R \left(1 + {c^2_0 k_z^2 \over \omega^2 - c^2_0 k_z^2 (1 +
D_2)}\right). \label{drhoapp} \eeq Substituting equations
(\ref{dbapp}) and (\ref{drhoapp}) into equation (\ref{frapp}), and
assuming $\beta \gg 1$ so that $|\omega^2| \ll k_z^2 c_0^2$, yields
\beq F_R = - k_R^2 \xi_R \left[c^2_0 \left(D_1 + {2D_2 \over 1 +
2D_2}\right) + v_{Az}^2\left(1 - {\omega^2 \over k_z^2 v_{Az}^2(1 +
2D_2)^2}\right)\right]. \label{fr} \eeq The MHD limit of equation
(\ref{fr}) can be obtained by setting $D_1 = D_2 = 0$ (see \S2.3).  In
this case $F_R = -\xi_R k_R^2 v_{Az}^2[1 + |\omega|^2/(k^2_z
v_{Az}^2)] \sim - \xi_R k_R^2 v_{Az}^2$.  Consider instead double
adiabatic theory, for which $D_1 = 1$ and $D_2 = 2$.  In this case
$F_R \sim -\xi_R k_R^2 c_0^2$; this is larger than the MHD pressure
force by a factor of $\sim \beta$.  Finally, for the full kinetic
problem we need to evaluate $D_1$ and $D_2$ using equation (\ref{D}).
Since the MRI has $|\omega| \lsim k_z v_A$ we can take $\xi \ll 1$ so
long as $\beta \gg 1$.  In this case $D_1 \approx D_2 \approx -i
\sqrt{\pi} \xi \sim -i \omega/k_z c_0$, so that $F_R \sim -k_R^2 \xi_R
c^2_0 (-i \omega / k_z c_0)$. To estimate the magnitude of $F_R$, note
that $|\omega| \sim k_z v_A$ in MHD, in which case $F_R \sim -\xi_R
k_R^2 v_{Az} c_0$.  This is $\sim \beta^{1/2}$ times larger than the
pressure force in MHD.  This large radial pressure gradient suppresses
the growth rates of the MRI, as seen in Figures 2-4.

\subsection{$B_\phi \ne 0; k_R = 0$}

In this case there is an azimuthal pressure force given by
(eq. [\ref{azi}]) \beq F_\phi \equiv -i k_z \sin\theta \cos\theta
{\dppar - \dpp \over \rho_0} = - i k_z c^2_0 \sin\theta \cos\theta
\left[D_2 {\drho \over \rho_0} - \left(D_1 + D_2\right){\delta B \over
B_0}\right], \label{fphiapp} \eeq where we have used equations
(\ref{dperp}) and (\ref{dpar}) to eliminate $\dpp$ and $\dppar$.  For
$\beta \gg 1$ the MRI is nearly incompressible and $\drho/\rho_0 \ll
\delta B/B_0$.\footnote{The calculation in Appendix A.1 shows this
explicitly: eqs. [\ref{drhoapp}] and [\ref{dbapp}] imply that
$\drho/\rho_0 \sim D_2 \delta B/B_0$ where $D_2 \sim \beta^{-1/2} \ll
1$.  The same qualitative result holds for the different geometry
considered here.}  We therefore neglect the $\drho/\rho_0$ term in
equation (\ref{fphiapp}).  Using $\delta v_\phi = {\partial
\xi_\phi/\partial t} - \xi_R d\Omega/d \ln R$ one can rewrite equation
(\ref{bpar}) for $\delta B$ in terms of $\xi_\phi$.  Again neglecting
$\delta \rho/\rho_0$ relative to the other terms, this yields \beq
{\delta B \over B_0} = - i k_z \sin\theta \cos \theta
\xi_\phi. \label{dbapp2} \eeq Substituting equation (\ref{dbapp2})
into equation (\ref{fphiapp}) yields \beq F_\phi = - \xi_\phi k^2_z
c^2_0 \sin^2\theta \cos^2\theta (D_2 + D_1). \label{fphi} \eeq In MHD,
$F_\phi = 0$, and magnetic tension, which $\sim - k_z^2 v_{Az}^2
\xi_\phi$ (see eq. [\ref{azi}]), plays the destabilizing role.  By
contrast, in kinetic theory the azimuthal pressure force is given by
$F_\phi \sim -\xi_\phi \sin^2\theta \cos^2\theta k_z^2 c_0^2 (-i
\omega/ k_z c_0)$.  For $B_\phi \sim B_z$, so that $\sin\theta \sim
\cos\theta \sim 1$, this is larger than the destabilizing azimuthal
tension force by a factor of $\sim \beta^{1/2}$.  This large
destabilizing azimuthal pressure force enhances the growth rates of
the MRI, as seen in Figures 2-4.

\end{appendix}

\newpage

\begin{figure}
\plotone{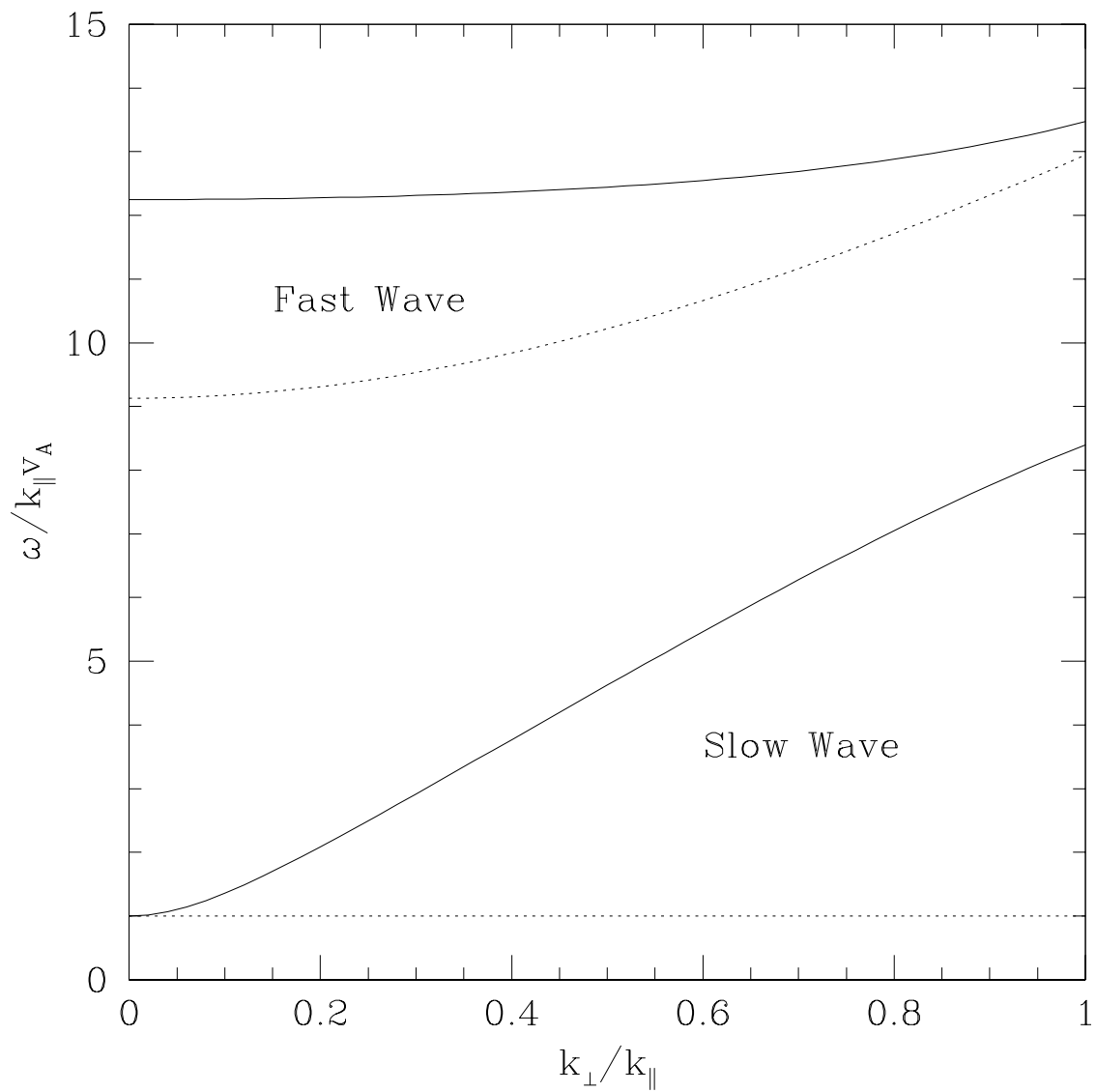}
\caption{The dispersion relation for the magnetosonic waves in MHD
(dotted lines) and in double adiabatic theory (solid lines), taking
$\beta = 100$.  In MHD, the slow wave dispersion relation is identical
to that of the Alfv\'en wave ($\omega = \kpar v_A$) while this is only
true for $\kp \ll \kpar$ in double adiabatic theory (see the
text for an explanation).}

\end{figure}

\begin{figure}
\plotone{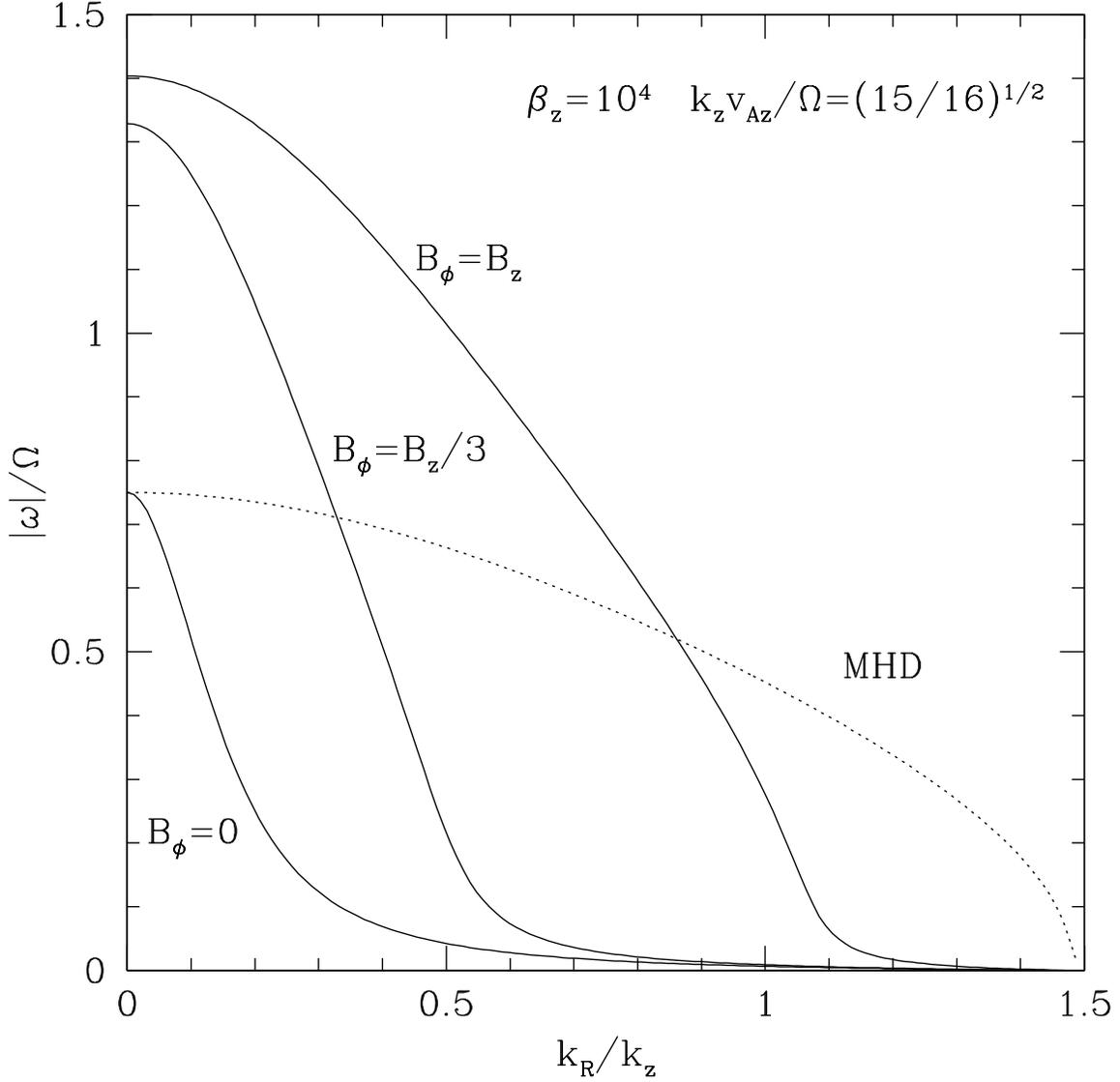}
\caption{The kinetic growth rates of the MRI for $\beta_z \equiv 8 \pi
p_0/B^2_z = 10^4$ and for different geometries of the seed magnetic
field.  The MHD results, which are independent of $B_\phi$, are shown
for comparison (dotted line).  The vertical wavenumber is taken to be
$k_z v_{Az}/\Omega = \sqrt{15/16}$, which is the fast growing mode in
MHD.}
\end{figure}

\begin{figure}
\plottwo{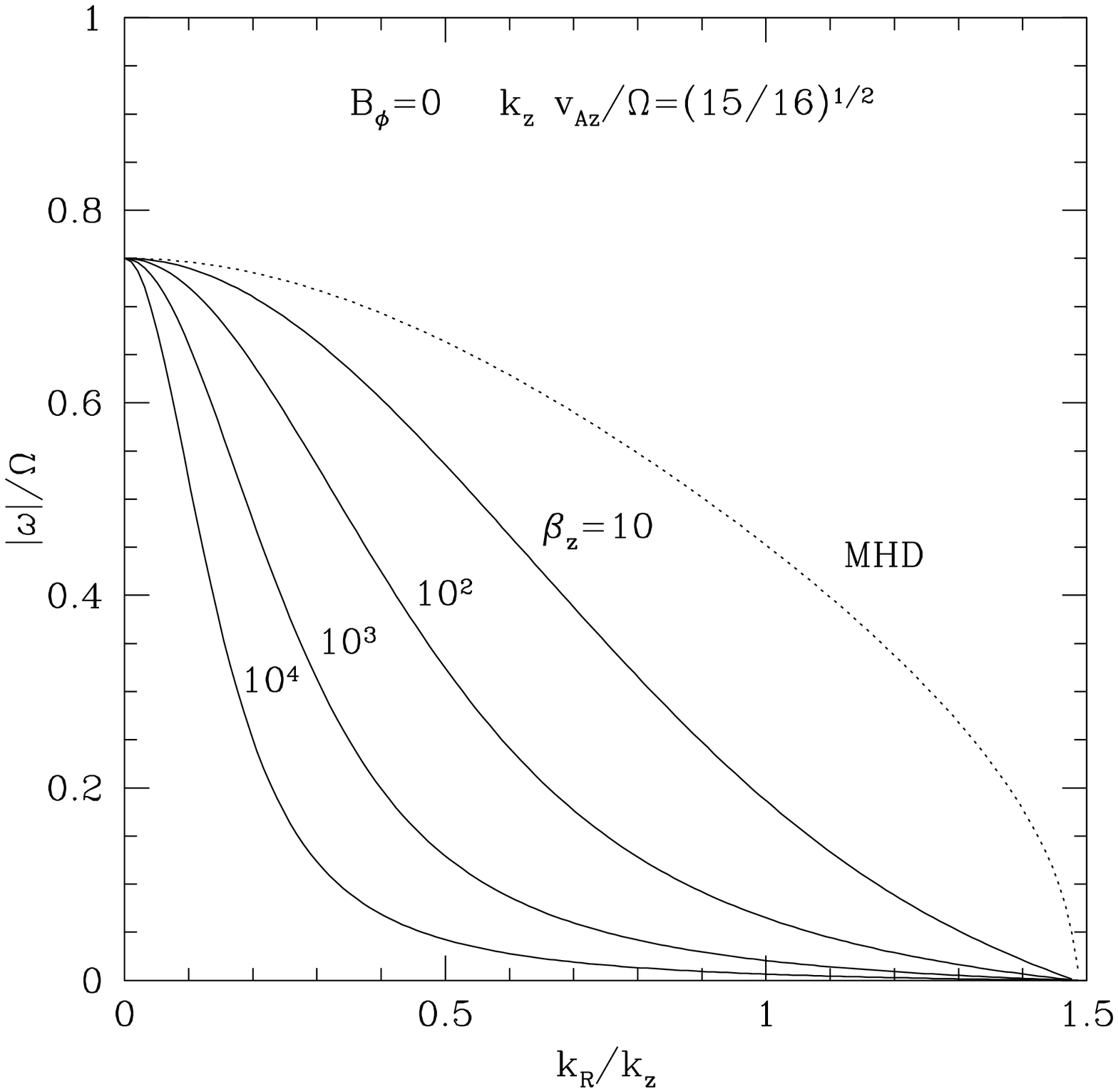}{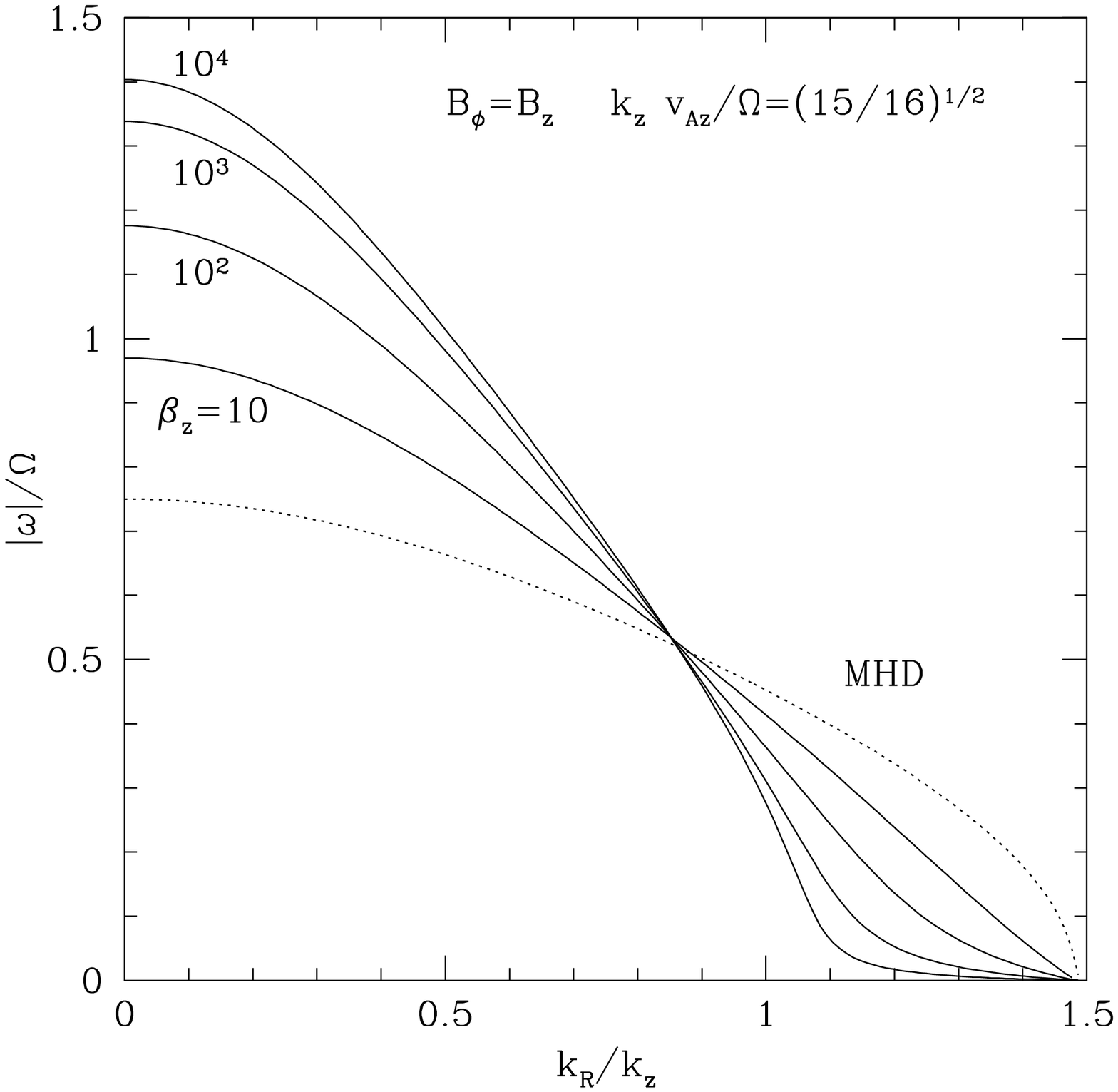}
\caption{The kinetic growth rates of the MRI for varying $\beta_z$
(solid lines).  The MHD results, which are nearly independent of
$\beta_z$, are shown for comparison (dotted line).  The vertical
wavenumber is taken to be $k_z v_{Az}/\Omega = \sqrt{15/16}$, which is
the fast growing mode in MHD.}
\end{figure}

\begin{figure}
\plottwo{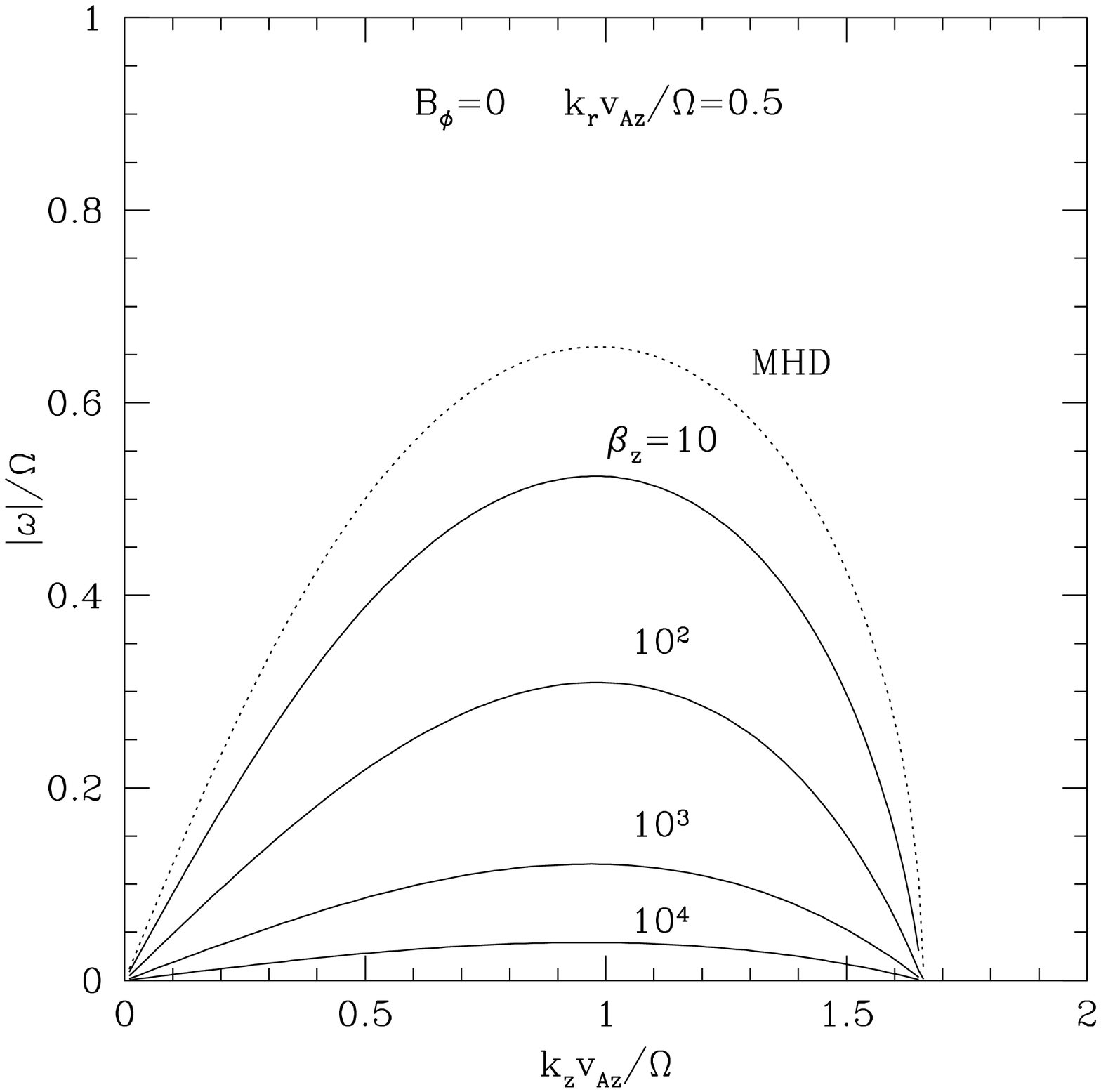}{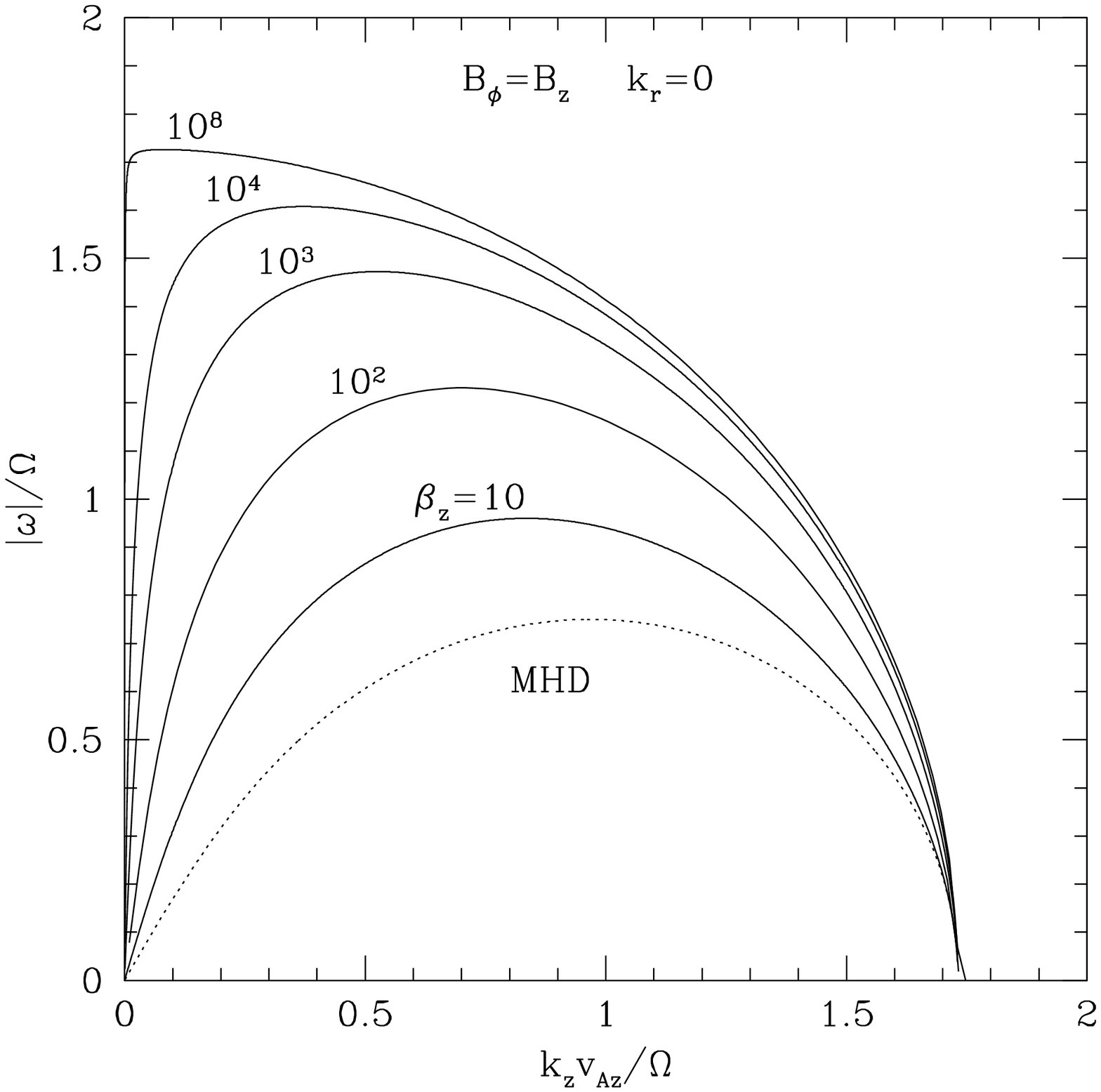}
\caption{The kinetic growth rates of the MRI as a function of $k_z$
for different $\beta_z$ (solid lines).  The corresponding MHD results
are shown by the dotted line.}
\end{figure}

\end{document}